\documentclass[a4paper,intlimits,superscriptaddress,8pt,floatfix]{iopart}
\expandafter\let\csname equation*\endcsname\relax
\expandafter\let\csname endequation*\endcsname\relax
\usepackage[tbtags]{amsmath}
\usepackage{amsthm,amssymb,dsfont,mathrsfs,bbm,xfrac,tabularx}
\usepackage{mathtools}
\usepackage{multirow}
\usepackage[cal=boondoxo]{mathalfa}
\DeclareMathAlphabet\mathrsfso{U}{rsfso}{m}{n}
\DeclareMathAlphabet\mathzapf{T1}{pzc}{mb}{it}
\usepackage[usenames,dvipsnames]{xcolor}
\usepackage[caption=false]{subfig}
\usepackage[english]{babel}
\usepackage{hyperref}
\hypersetup{pdfauthor={Sicuro Zdeborova},pdftitle={Planted k-factor},%
            colorlinks, linktocpage=true, pdfstartpage=3, pdfstartview=FitV,%
    breaklinks=true, pdfpagemode=UseNone, pageanchor=true, pdfpagemode=UseOutlines,%
    plainpages=false, bookmarksnumbered, bookmarksopen=true, bookmarksopenlevel=1,%
    hypertexnames=true, pdfhighlight=/O,%
    urlcolor=orange, linkcolor=blue, citecolor=red, 
        }
\usepackage{tikz}
\usetikzlibrary{patterns,decorations,arrows,shapes.geometric,arrows,mindmap,backgrounds,positioning,positioning,fit,backgrounds,positioning,arrows,matrix,calc}
\usetikzlibrary{shapes,backgrounds,mindmap,shadows,shapes.geometric,topaths,snakes,arrows,pgfplots.groupplots,fadings,calc,decorations.markings,positioning,chains,fit}

\newcommand{\dd}{{\rm d}}

\DeclareMathOperator{\Prob}{{\mathbb P}}
\DeclareMathOperator{\supp}{supp}

\newcommand{\mG}{{\mathrsfso G}}
\newcommand{\mF}{{\mathrsfso F}}
\newcommand{\mE}{{\mathcal E}}
\newcommand{\mV}{{\mathcal V}}

\usepackage{libertine}\usepackage[libertine]{newtxmath}

\allowdisplaybreaks

\begin{document}
\title[The planted $k$-factor problem]{The planted \boldmath{$k$}-factor problem}
\author{Gabriele Sicuro}
\address{IdePHICS laboratory, EPFL, 
1015 Lausanne, Switzerland}
\address{Department of Mathematics, King's College London, London WC2R 2LS, UK}
\author{Lenka Zdeborov\'a}
\address{SPOC laboratory, EPFL,
1015 Lausanne, Switzerland}
\date{\today}
\begin{abstract}
We consider the problem of recovering an unknown $k$-factor, hidden in a weighted random graph. For $k=1$ this is the planted matching problem, while the $k=2$ case is closely related to the planted travelling salesman problem. The inference problem is solved by exploiting the information arising from the use of two different distributions for the weights on the edges inside and outside the planted sub-graph. We argue that, in the large size limit, a phase transition can appear between a full and a partial recovery phase as function of the signal-to-noise ratio. We give a criterion for the location of the transition.
\end{abstract}

\section{Introduction}

The problem of recovering a hidden structure (the signal) in a graph on the basis of the observation of its edges and the weights on edges and vertices appears in many, diverse contexts. Community detection~\cite{decelle2011asymptotic}, group testing~\cite{mezard2008group}, certain types of  error correcting codes~\cite{richardson2008modern}, particle tracking \cite{Chertkov2010} are some example of statistical inference problems formulated on graphs in which some underlying pattern has to be identified. The feasibility of the hidden structure recovery depends, of course, on the amount of ``noise'' in the problem. It turns out that, in the limit of large system sizes, sharp recovery thresholds with respect to the signal-to-noise ratio can appear. These sharp thresholds separate no recovery phases (in which no information on the signal is accessible), partial recovery phases (in which an output correlated with the signal can be obtained) and full recovery phases (in which the signal is recovered with a vanishing error). 

These algorithmic transitions, analogous to phase transitions in physics, can be of different orders (depending on the number of derivatives of the order parameter that exist). The order of the transition seems to have important algorithmic implications. First order transitions, in particular, have been related to the presence of a computationally hard phases \cite{decelle2011asymptotic}. In the Stochastic Block Model (SBM) on sparse graphs, for example, a phase transition between a no-recovery phase and a partial recovery phase is found~\cite{decelle2011asymptotic}. 
In \cite{Abbe2015,Abbe2016} it has been shown that a partial to full recovery transition can also appear in the same model if denser topologies are considered. A partial to full recovery first order transition appears in low-density-parity-check error correcting codes~\cite{richardson2008modern}, where the target is to correctly recover a code-word. Somewhat uncommonly, a partial to full recovery infinite order phase transition has been recently found in the planted matching problem, in which a hidden matching has to be detected in a weighted random graph \cite{Moharrami2019,Semerjian2020,Ding2021}. This is a new type of a phase transition found in inference problems that has been put in relation with the presence of instabilities of the belief propagation fixed points. It is thus of interest to investigate whether it appears in other problems than the planted matching. 

In this paper we study a generalization of the planted matching problem  -- the so-called planted $k$-factor problem. A $k$-factor of a graph $\mG$ is a spanning subgraph with fixed degree $k$, i.e., a spanning $k$-regular graph. In this problem, the (weighted) $k$-regular graph is hidden (planted) by adding new weighted edges to it. The weights of the planted and non-planted edges are random quantities with different distributions. The goal is to recover the planted $k$-factor knowing $k$ and the two weight distributions. 

In analogy with percolation being continuous while the appearance of a $k$-core is a discontinuous (first order) phase transition \cite{stauffer2018introduction}, we may have anticipated that the transition in the $k$-factor problem will be of a different type than in the planted matching. We will show instead that the planted $k$-factor problem manifests a partial--full recovery continuous phase transition akin the one in the planted matching, and we will give a criterion for the threshold between the two phases. 

The planted $k$-factor problem is related to many interesting inference problems. For example, it shares some similarity with the problem of structure-detection in small-world graphs \cite{Cai2017}. In the latter case, a $k$-regular ring lattice is hidden in a (unweighted) random graph by a rewiring procedure. Full recovery is possible depending on the parameters of the rewiring process.

The planted $1$-factor problem corresponds to the aforementioned planted matching problem, introduced in~\cite{Chertkov2010} as a model for particle tracking. In this model, a weighted perfect matching, hidden in a graph, has to be recovered. In Ref.~\cite{Chertkov2010} the problem was studied numerically for a particular case of the distribution of weights. The results suggested the existence of a phase transition between a full recovery phase and a partial recovery phase. More recently, Moharrami and coworkers \cite{Moharrami2019} rigorously analysed another special setting of the problem assuming that $\mG$ is the complete graph and the weights are exponentially distributed, and proved the existence of a partial/full recovery phase transition. The results of Ref.~\cite{Moharrami2019} have been generalized to sparse graphs and general weight distributions in Ref.~\cite{Semerjian2020}, via heuristic arguments based on a correspondence between the problem and branching random walk processes. A proof of the results in \cite{Semerjian2020} has been recently given in \cite{Ding2021}.

The planted $2$-factor problem is a relaxation of the planted travelling salesman problem \cite{Bagaria2018}. In this problem, a unique, hidden Hamiltonian cycle visiting (exactly once) all vertices of a graph has to be recovered. In the planted $2$-factor problem, instead, the planted subgraph is more generally given by a set of cycles. Solving the $2$-factor problem can be, however, informative on a hidden Hamiltonian cycle. In Ref.~\cite{Bagaria2018} the Hamiltonian cycle recovery problem has been studied on the complete graph. Therein, a sufficient condition for the full recovery of the planted Hamiltonian cycle has been derived. Moreover, it is proved therein that, within the full recovery phase, the solution obtained searching for a $2$-factor coincides (with high probability and in the large size limit) with the hidden Hamiltonian cycle. 

In this paper, we generalize the available results for the planted $1$-factor problem and planted $2$-factor problem to the general planted $k$-factor problem with arbitrary distributions for the edge weights and including sparse graphs in the analysis. Unlike Ref.~\cite{Cai2017}, we will assume no prior knowledge on the structure of the planted $k$-factor (except for the degree of its vertices). Our approach, based on the standard cavity method and the corresponding belief propagation equations~\cite{mezard2009information}, allows us to obtain, at the level of rigor of theoretical physics, a criterion for the full recovery of the planted subgraph in the large size limit of the problem. The threshold criterion is derived studying the recursive distributional equations corresponding to the cavity equations. It turns out that a ``drift'' in their solutions can appear under iteration. If this is the case, the full-recovery solution is the only stable one, and full recovery is possible. The study of such drift has been tackled in analogy with the analysis in Ref.~\cite{Semerjian2020} for the $k=1$ case, and with the phenomenon of front propagation for reaction-diffusion equations \cite{Kingman1975,Biggins1977,Brunet1997,Majumdar2000,Ebert2000}. We give an explicit criterion for the threshold between a partial recovery phase and a full recovery phase of the planted $k$-factor. Our results recover, as special cases, the ones obtained in Refs.~\cite{Moharrami2019,Semerjian2020,Ding2021} for the planted $1$-factor. In the limit of dense graphs, they provide a sharper characterization of the phase transition for $k=2$ with respect to the analysis in \cite{Bagaria2018}, as discussed in more detail in Section \ref{sec:fully_connected}. 

The rest of the paper is organized as follows. In Section~\ref{sec:definitions} we define the problem under study and introduce two adopted statistical estimators, namely the block Maximum A Posteriori (MAP) and the symbol MAP. In Section \ref{sec:BP} we present the belief propagation equations for the solution of the problem and their corresponding probabilistic description. In Section~\ref{sec:Exp} we numerically study a specific case, observing that a transition between a full recovery and a partial recovery phase can appear as function of the parameters of the problem. In Section~\ref{sec:criterion} we give a heuristic derivation of the criterion for the location of the transition for arbitrary weight distributions for the block MAP estimator, Eq.~\eqref{condizione2}, which is the main result of the paper. In Section~\ref{sec:app} we compare our theoretical predictions with the numerical results obtained for different variants of the problem, including the case considered in Section~\ref{sec:Exp} and the Hamiltonian cycle recovery problem considered in Ref.~\cite{Bagaria2018}. Finally, conclusions and perspectives for future work are given in Section \ref{sec:conclusions}.

\section{Definitions}\label{sec:definitions}

\subsection{Planting a $k$-factor}
Let us assume that an integer $k\in\mathds N$ and two probability densities $p$ and $\hat p$ on the real line are given. We will focus on an ensemble of weighted simple graphs denoted $\mG_0=(\mV_0, \mE_0,\underline{w})$, containing by construction a planted $k$-factor to be recovered. Here $\mV_0 = \{1,\dots,N\}$ is the set of $N\in\mathds N$ vertices such that $kN$ is even, and $\mE_0$ is the set of edges (unordered pairs of distinct vertices of $\mV_0$). A weight $w_e\in\mathds R$ is associated to each edge $e$ of the graph, so that $\underline w=\{w_e \colon e \in \mE_0\}$ is the set of such weights. We introduce a probability measure over the set of weighted graphs by means of the following generation steps of $\mG_0$ (see Fig.~\ref{figA}).
\begin{enumerate}
\item One first constructs a $k$-regular graph having vertex set $\mV_0$. The graph is chosen uniformly among all possible $k$-regular graph with $N$ vertices \cite{Bollobas2001book}. This can be achieved using fast algorithms available in the literature \cite{Gao2017}. The obtained graph has $\frac{1}{2}kN$ edges and edge-set $\mF^*_k$. A random weight $w_e$, generated independently of all the others with density $\hat p$, is associated to each edge $e\in\mF^*_k$.
\item Given a pair of vertices that are not neighbours in $\mF^*_k$, a link is added between them with probability $cN^{-1}$. Let $\mE_0$ be the final set of edges of the obtained graph. A weight $w_e$, independently generated from all the others with distribution $p$, is assigned to each $e \in \mE_0 \setminus \mF^*_k$.
\end{enumerate}
We shall call planted (resp.~non-planted) edges those in $\mF^*_k$ (resp.~in $\mE_0 \setminus \mF^*_k$). The parameters of this ensemble of weighted random graphs are thus the integers $N$ and $k$, the parameter $c$ controlling the density of non-planted edges, and the two distributions $\hat{p}$ and $p$ for the generation of the weights of the planted and non-planted edges respectively. Given $\mF^*_k$, the probability to generate a graph $\mG_0$ is therefore
\begin{equation} \label{eq_proba_direct}
\mathbb P(\mG_0 | \mF^*_k)= \mathbb{I}(\mF^*_k \subseteq \mE_0)\prod_{e \in \mF^*_k} \hat p(w_e) \prod_{\mathclap{e \in \mE_0 \setminus \mF^*_k}} p(w_e) \left(\frac{c}{N} \right)^{|\mE_0|-k\frac{N}{2}} \left(1-\frac{c}{N} \right)^{\binom{N}{2} - |\mE_0|} ,
\end{equation}
where here and in the following $\mathbb{I}(A)$ denotes the indicator function of the event $A$. The edges in $\mE_0\setminus\mF^*_k$ form essentially an Erd\H os-R\'enyi random graph of average degree $c$. The edge-set $\mF^*_k$, on the other hand, is a $k$-factor of $\mG_0$ by construction, i.e., a spanning subgraph of $\mG_0$ in which all the vertices have the same degree $k$. The resulting graph has average coordination $c+k$. Note that, for $k=1$, $\mF^*_1$ is a matching on $\mG_0$, and the introduced ensemble of graphs coincides with the one studied in Ref.~\cite{Semerjian2020} for the planted matching problem.

\subsection{The inference problem}
\label{sec_def_inference}
Given a graph $\mG_0$ in the ensemble described above, we wonder if it si possible to infer the $k$-factor $\mF^*_k$ hidden in it. We assume that the generation rules are known, alongside with $k$, $c$, $p$ and $\hat p$. All the exploitable information is contained in the posterior probability $\mathbb P(\mF_k|\mG_0)$ that a certain $k$-factor $\mF_k$ in $\mG_0$ is the planted $k$-factor $\mF_k^*$ From Bayes theorem we obtain the following expression for the posterior:
\begin{equation}
\mathbb P(\mF_k| \mG_0 ) \propto \mathbb{I}(\mF_k\text{  is a $k$-factor})\prod_{e \in \mF_k} \hat p(w_e) \prod_{\mathclap{e \in \mE_0 \setminus \mF_k}} p(w_e) \mathbb{I}(\mF_k \subseteq  \mE_0 ) ,
\end{equation}
where the symbol $\propto$ hides a normalization constant independent of $\mF_k$. To parametrize the probability measure above, it is convenient to introduce, for each edge $e$, the binary variable $m_e\in\{0,1\}$, so that
$\underline{m}=\{m_e=\mathbb I(e\in \mF_k) \colon e \in \mE_0\} \in \{0,1\}^{|\mE_0|}$, and rewrite the posterior as
\begin{equation}\label{post}
\mathbb P( \underline{m}| \mG_0 ) \propto \prod_{e \in \mE_0 } \left( \frac{\hat p(w_e)}{p(w_e)}  \right)^{m_e}  \prod_{i=1}^N \mathbb{I}\left(\sum_{e \in \partial i} m_e = k \right) \ ,
\end{equation}
where $\partial i$ denotes the set of edges incident to the vertex~$i$. We want to compute an estimator $\hat \mF_k$ that is ``close'' to the hidden $k$-factor $\mF_k$. The estimator is associated to the set of binary variables $\hat{\underline{m}}$ that encodes the set of edges in $\hat\mF_k$. With a slight abuse of notations, in the following we identify a set of edges $\mF_k$ with its corresponding ${\underline{m}}$. We will denote $\underline m^*$ the set of variables associated to $\mF_k^*$ and $\hat{\underline m}$ the set of variables associated to an estimator $\hat\mF_k$. In this paper, we will quantify the distance between an estimator $\hat\mF_k$ and the true planted $k$-factor $\mF_k^*$ in terms of the cardinality of the symmetric difference $\mF_k^* \triangle \hat{\mF}_k$ between $\mF_k^*$ and $\hat\mF_k$,
\begin{equation}
\varrho(\mF_k^*,\hat{\mF}_k) \coloneqq \frac{|\mF_k^* \triangle \hat{\mF}_k|}{2|\mF_k^*|}= \frac{1}{kN} \sum_{e \in \mE_0 }\mathbb{I}(\hat{m}_e\neq m^*_e),
\label{eq_def_rho}
\end{equation}
or equivalently the Hamming distance between the binary string $\underline{m}^*$ encoding $\mF_k^*$ and the binary string $\hat{\underline{m}}$ encoding $\hat{\mF}_k$. We will consider two `Maximal A Posteriori' (MAP) estimators, each one minimizing a ``measure of distance'' with the planted $k$-factor.
\begin{itemize}
 \item A first possibility is to choose as estimator the $k$-factor that minimizes the probability $\mathbb{P}( \mF_k^* \neq \hat{\mF}_k)$ over all realizations of the problem,
\begin{equation}\label{eq_bMAP}
\hat{\mF}^{\rm (b)}_k = \underset{\underline{m}}{\rm argmax}\ \mathbb P( \underline{m}| \mG_0 ).
\end{equation}
This estimator is usually called `block MAP' \cite{richardson2008modern}.
\item A different estimator, called `symbol MAP' and denoted in the following $\hat{\mF}_k^{\rm (s)}$, is obtained requiring that the distance to be minimized is precisely the error in Eq.~\eqref{eq_def_rho}. In this case, for each edge $e \in \mE_0$, we choose
\begin{equation} \label{eq_sMAP}
\hat{m}_e= \underset{m_e}{\rm argmax}\ \mathbb P_e(m_e | \mG_0 ),
\end{equation}
with $\mathbb P_e$ the marginal of the posterior probability \eqref{post} for the edge $e$. Observe, however, that this estimator is not necessarily a $k$-factor.
\end{itemize}
In the following, we will discuss both the estimators defined above, in the thermodynamic limit $N \to \infty$, as a function of the parameters of the model. As in the planted matching problem \cite{Chertkov2010,Moharrami2019,Semerjian2020}, the possibility of identifying the planted edges will depend on the similarity between the distribution $p$ and the distribution $\hat p$, and on the parameter $c$, that corresponds itself to a noise level expressing the number of confusing non-planted edges introduced in the graph.

\section{Cavity equations}\label{sec:BP}
\begin{figure*}
\subfloat[\label{figA} $\mG_0$]{\includegraphics[width=0.24\textwidth]{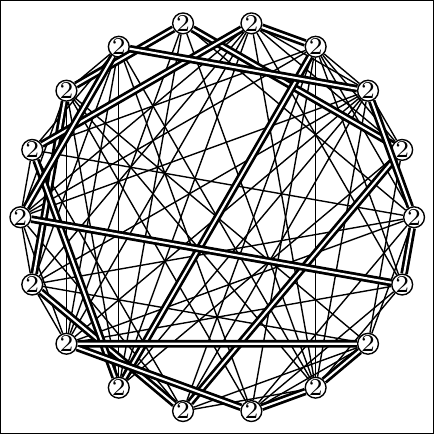}}\hfill
\subfloat[\label{figB}]{\includegraphics[width=0.24\textwidth]{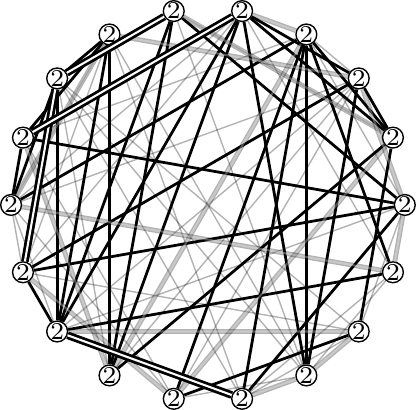}} \hfill
\subfloat[\label{figC} $\mG$]{\includegraphics[width=0.24\textwidth]{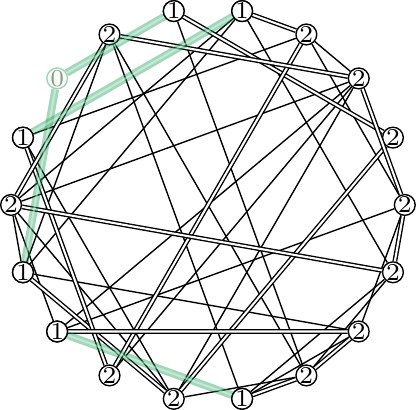}} 
\caption{Pictorial representation of the preliminary pruning in a planted $2$-factor problem. (a) A $2$-regular graph on a vertex set of $N$ vertices is generated (thick double lines; in the picture, $N=18$), with edge weights distributed as $\hat p$. Random edges are added with some probability $\sfrac{c}{N}$ (thin lines), with edge weights distributed as $p$. The obtained graph is $\mG_0$. A capacity $k$ is assigned to each vertex. (b) Edges in $\supp(\hat p)\setminus\Gamma$ (double lines) can be identified as planted and removed, decreasing the capacity of the endpoints by $1$. Edges in $\supp(p)\setminus\Gamma$ (thick single lines) can be identified and removed. Vertices with zero capacity can be removed. (c) We call $\mG$ the obtained pruned graph, and we call $\mF_k^{(0)}$ the set of identified edges by means of this first pruning process (green). 
}
\end{figure*}

\subsection{Pruning the graph}\label{sec:pruning}
To efficiently solve the problem, and possibly reduce the size of the input, it is convenient to proceed with a preliminary ``pruning'' of the graph. Before proceeding further, let us assign a `capacity variable' $\kappa_i=k$ to each vertex $i$. The capacity of each node $i$ will take into account the number of unidentified planted edges incident at $i$. Let 
\begin{subequations}
\begin{align}
\supp(\hat p)&\coloneqq\{w\in\mathds{R}\colon \hat p(w)>0\},\\
\supp(p)&\coloneqq\{w\in\mathds{R}\colon p(w)>0\},
\end{align}
be the support of $\hat p$ and $p$ respectively, and let
\begin{equation}
\Gamma\coloneqq\supp(p)\cap\supp(\hat p) \label{def:gamma}
\end{equation}
be the intersection of these supports. We suppose that $\Gamma\neq\emptyset$ (the inference problem is otherwise trivial). 

If an edge $e$ bears a weight $w_e\in\supp(p)\setminus\Gamma$, then it can be immediately identified as `non-planted', and removed from $\mG_0$. This event will happen with probability $1-\mu$, where
\begin{equation}
\mu\coloneqq\int_{\Gamma}p(w)\dd w.
\end{equation}
On the other hand, the set of edges
\begin{equation}
{\mF}_k^{(0)}\coloneqq \left\{e\in\mE_0\colon w_e \in \supp(\hat p) \setminus \Gamma \right\}
\end{equation}
surely belongs to the planted configuration, ${\mF}_k^{(0)}\subseteq\mF_k^*$. These edges can be correctly classified as `planted' with no algorithmic effort, except for the inspection of their weights. A planted edge $e$ can be therefore identified, solely on the basis of the value of its weight, with probability $1-\hat\mu$, where
\begin{equation}
\hat\mu\coloneqq\int_{\Gamma}\hat p(w)\dd w.
\end{equation}
\end{subequations}
We can remove from the graph the identified planted edges, see Fig.~\ref{figB}. We must take care, however, of reducing at the same time by $1$ the capacity of the endpoints of a planted edge that is removed. At the end of this process, the capacity of a generic vertex $i$ is $0\leq \kappa_i\leq k$, see Fig.~\ref{figC}. For large $N$, the capacity of the vertices after this pruning has binomial distribution ${\rm Bin}(k,\hat\mu)$. In particular, $(1-\hat\mu)^k N$ vertices have zero capacity, meaning that all their incident planted edges have been identified. These vertices can also be removed, alongside with all their remaining (non-planted) incident edges. 

In the resulting pruned graph, a vertex has $\mathsf K$ incident planted edges and $\mathsf Z$ non-planted edges, where $\mathsf K$ and $\mathsf Z$ are two random variables having distribution
\begin{subequations}
\begin{align}
\Prob[\mathsf K=\kappa]&=\binom{k}{\kappa}\frac{\hat\mu^\kappa(1-\hat\mu)^{k-\kappa}}{\hat\mu_k},\quad \kappa=1,\dots,k,\label{distK}\\
\Prob[\mathsf Z=z]&=\gamma^{z}\frac{\e^{-\gamma}}{z!},\quad z\in\mathds N_0,\quad \gamma\coloneqq c\mu\hat\mu_k,
\end{align}
\end{subequations}
where $\hat\mu_k\coloneqq 1-(1-\hat\mu)^k$, so that the pruned graph has average degree $\mathbb E[\mathsf K+\mathsf Z]=\gamma+k\hat\mu\hat\mu_k^{-1}$. The distributions of the weights of the surviving edges are obtained from the original ones conditioning the weights to be in $\Gamma$, i.e.,
\begin{subequations}\label{PPhat}
\begin{align}
P(w)&\coloneqq \frac{p(w)}{\mu}\mathbb I(w\in\Gamma), \\
\hat P(w) &\coloneqq \frac{\hat p( w)}{\hat\mu}\mathbb I( w\in\Gamma),
\end{align}
\end{subequations}
for the non-planted and planted edges, respectively. We will denote $\mG =(\mV, \mE, \underline{w})$ the pruned graph, with $\mV\subseteq\mV_0$ and $\mE \subseteq \mE_0$ the new vertex and edge-sets. For large $N$, $\mV$ has cardinality $\hat N\coloneqq \hat\mu_kN$, each vertex $i\in\mV$ having capacity $1\leq \kappa_i\leq k$ distributed as $\mathsf K$. The graph will have a total of $\frac{1}{2}k\hat\mu N$ surviving planted edges and $\frac{1}{2}\gamma\hat\mu_k N$ surviving non-planted edges.

\subsection{The Belief Propagation equations}

To write down a belief propagation algorithm for the planted $k$-factor problem, we start from the probability distribution over the configurations $\underline{m}=\{m_e : e \in \mE\} \in \{0,1\}^{|\mE|}$ on the edges of the pruned graph $\mG$,
\begin{equation}\label{lbeta}
\nu(\underline{m})\propto \exp\left(-\beta\sum_{e \in\mE} m_e \omega_e\right)\prod_{i \in\mV}\mathbb I\left(\sum_{e\in \partial i}m_e=\kappa_i\right),
\end{equation}
where $\beta>0$ and $\omega_e = \omega(w_e)$, with
\begin{equation}\label{omega} 
\omega(w)\coloneqq -\ln\frac{\hat P(w)}{P(w)}.
\end{equation}
The quantities $\omega_e$ play the role of effective weights on the edges of the graph. The introduction of $\beta$ is convenient because the measure in Eq.~\eqref{lbeta} coincides, for $\beta=1$, with the posterior defined in Eq.~\eqref{post}. On the other hand, for $\beta \to \infty$ the measure concentrates on the configurations maximizing the posterior, i.e., on the block MAP. Eq.~\eqref{lbeta} can also be associated to a graphical model. In particular, we can associate a variable vertex $m_e$ (\tikz{\node[shape=circle,draw,inner sep=2pt] (e) {};}) to each edge $e$ of $\mG$. We also introduce two types of interaction vertices. A first type of interaction vertex (\tikz{\node[fill=black,shape=rectangle,draw,inner sep=2pt] (b4) {};}) is associated to each $i \in \mV$, and linked to all variable vertices $m_e$ such that $e\in\partial i$. Such vertex imposes that $\kappa_i$ variables $m_e$, $e\in\partial i$, are equal to 1. A second interaction vertex expresses the contribution $\e^{-\beta m_e \omega_e}$ (\tikz{\node[fill=gray!60, shape=rectangle, draw,inner sep=2pt] (B4) {};}) for each $e$, and it is linked to the variable vertex $m_e$. {Pictorially,
\newdimen\nodeSize
\nodeSize=4mm
\newdimen\nodeDist
\nodeDist=6mm
\tikzset{position/.style args={#1:#2 from #3}{at=(#3.#1), anchor=#1+180, shift=(#1:#2)}}
\[\underbrace{\begin{gathered}
\begin{tikzpicture}
\node[fill=black,shape=circle,draw,inner sep=1pt] (0) at (0,0) {};
\node[fill=black,shape=circle,position=120:{0.35*\nodeDist} from 0,draw,inner sep=1pt] (a2) {};
\node[draw=none,position=120:{0.35*\nodeDist} from a2,inner sep=1pt] (a2a) {};
\node[draw=none,position=180:{0.35*\nodeDist} from a2,inner sep=1pt] (a2b) {};
\node[draw=none,position=60:{0.35*\nodeDist} from a2,inner sep=1pt] (a2c) {};
\node[fill=black,shape=circle,position=240:{0.35*\nodeDist} from 0,draw,inner sep=1pt] (a3) {};
\node[draw=none,position=-120:{0.35*\nodeDist} from a3,inner sep=1pt] (a3a) {};
\node[draw=none,position=-180:{0.35*\nodeDist} from a3,inner sep=1pt] (a3b) {};
\draw[thin,gray] (a2a) -- (a2) -- (a2b) -- (a2) -- (a2c);
\draw[thin,gray] (a3a) -- (a3) -- (a3b);
\draw[thin] (a3) -- (0) -- (a2);
\node[fill=black,shape=circle, position=0:10mm from 0,draw,inner sep=1pt] (i) {};
\node[fill=black,shape=circle,position=60:{0.35*\nodeDist} from i,draw,inner sep=1pt] (b4) {};
\node[fill=black,shape=circle,position=-60:{0.35*\nodeDist} from i,draw,inner sep=1pt] (b5) {};
\draw[thin] (b4) -- (i) -- (b5);
\node[draw=none,position=0:{0.4*\nodeDist} from b4,inner sep=1pt] (b4a) {};
\node[draw=none,position=60:{0.4*\nodeDist} from b4,inner sep=1pt] (b4b) {};
\node[draw=none,position=10:{0.4*\nodeDist} from b5,inner sep=1pt] (b5a) {};
\node[draw=none,position=-40:{0.4*\nodeDist} from b5,inner sep=1pt] (b5b) {};
\node[draw=none,position=-90:{0.4*\nodeDist} from b5,inner sep=1pt] (b5c) {};
\node[draw=none,position=-140:{0.4*\nodeDist} from b5,inner sep=1pt] (b5d) {};
\draw[thin,gray] (b4a) -- (b4) -- (b4b);
\draw[thin,gray] (b5a) -- (b5) -- (b5b) -- (b5) -- (b5c) -- (b5) -- (b5d);
\draw[thin] (0) -- (i);
\end{tikzpicture}
\end{gathered}}_{\mG}\Longrightarrow
\begin{gathered}
\begin{tikzpicture}
\node[fill=black,shape=rectangle,draw,inner sep=2pt] (0) at (0,0) {};
\node[shape=circle,draw,inner sep=2pt, position=0:5mm from 0] (e) {};
\node[ fill=gray!60, shape=rectangle, position=90:{0.2\nodeDist} from e, draw,inner sep=2pt] (A) {};
\node[fill=black,shape=rectangle,position=120:{0.7*\nodeDist} from 0,draw,inner sep=2pt] (a2) {};
\node[shape=circle, position=120:{0.2\nodeDist} from 0,draw,inner sep=2pt] (2) {};
\node[ fill=gray!60, shape=rectangle, position=30:{0.2\nodeDist} from 2, draw,inner sep=2pt] (A2) {};
\node[fill=black,shape=rectangle,position=240:{0.7*\nodeDist} from 0,draw,inner sep=2pt] (a3) {};
\node[shape=circle, position=240:{0.2\nodeDist} from 0,draw,inner sep=2pt] (3) {};
\node[ fill=gray!60, shape=rectangle, position=-30:{0.2\nodeDist} from 3, draw,inner sep=2pt] (A3) {};
\draw[thin] (e) -- (0) -- (2) -- (a2) -- (2) -- (A2);
\draw[thin] (0) -- (3) -- (a3) -- (3) -- (A3);
\node[fill=black, position=0:5mm from e,draw,inner sep=2pt] (i) {};
\node[fill=black,shape=rectangle,position=60:{0.7*\nodeDist} from i,draw,inner sep=2pt] (b4) {};
\node[shape=circle, position=60:{0.2\nodeDist} from i,draw,inner sep=2pt] (4) {};
\node[ fill=gray!60, shape=rectangle, position=150:{0.2\nodeDist} from 4, draw,inner sep=2pt] (B4) {};
\node[fill=black,shape=rectangle,position=-60:{0.7*\nodeDist} from i,draw,inner sep=2pt] (b5) {};
\node[shape=circle, position=-60:{0.2*\nodeDist} from i,draw,inner sep=2pt] (5) {};\node[ fill=gray!60, shape=rectangle, position=-150:{0.25\nodeDist} from 5, draw,inner sep=2pt] (B5) {};
\draw[thin] (i) -- (4) -- (b4) -- (4) -- (B4);
\draw[thin] (i) -- (e) -- (A);
\draw[thin] (i) -- (5) -- (b5) -- (5) -- (B5);
\node[draw=none,position=120:{0.35*\nodeDist} from a2,inner sep=1pt] (a2a) {};
\node[draw=none,position=180:{0.35*\nodeDist} from a2,inner sep=1pt] (a2b) {};
\node[draw=none,position=60:{0.35*\nodeDist} from a2,inner sep=1pt] (a2c) {};
\node[draw=none,position=-120:{0.35*\nodeDist} from a3,inner sep=1pt] (a3a) {};
\node[draw=none,position=-180:{0.35*\nodeDist} from a3,inner sep=1pt] (a3b) {};
\draw[thin,gray] (a2a) -- (a2) -- (a2b) -- (a2) -- (a2c);
\draw[thin,gray] (a3a) -- (a3) -- (a3b);
\node[draw=none,position=0:{0.4*\nodeDist} from b4,inner sep=1pt] (b4a) {};
\node[draw=none,position=60:{0.4*\nodeDist} from b4,inner sep=1pt] (b4b) {};
\node[draw=none,position=10:{0.4*\nodeDist} from b5,inner sep=1pt] (b5a) {};
\node[draw=none,position=-40:{0.4*\nodeDist} from b5,inner sep=1pt] (b5b) {};
\node[draw=none,position=-90:{0.4*\nodeDist} from b5,inner sep=1pt] (b5c) {};
\node[draw=none,position=-140:{0.4*\nodeDist} from b5,inner sep=1pt] (b5d) {};
\draw[thin,gray] (b4a) -- (b4) -- (b4b);
\draw[thin,gray] (b5a) -- (b5) -- (b5b) -- (b5) -- (b5c) -- (b5) -- (b5d);
\end{tikzpicture}
\end{gathered}\]}
The Belief Propagation (BP) algorithm \cite{mezard2009information} provides a recipe for the computation of the marginals of $\nu$ on such factor graph. The idea is to approximate the marginal $\nu_e(m)$ for the edge $e=(i,j)$ as
\[\nu_e(m)\coloneqq \sum_{\{m_{\tilde e}\}_{\tilde e\neq e}}\nu(\underline{m})\simeq \nu_{i\to e}(m)\nu_{j\to e}(m)e^{-\beta m\omega_e},\]
where $\nu_{i\to e}(m)$ (respectively $\nu_{i\to e}(m)$) mimics a marginal probability in graphical models in which $j$ (respectively $i$) is absent. Such factorization is exact on infinite trees. The algorithm goal is the computation of such messages, and it is conjectured to be exact in the large size limit for sparse random graphs. The messages obey the following equations (one for each directed edge of the graph),
\begin{equation}\label{messaggio}
 \nu_{i\to e}(m)\propto
 \sum_{\mathclap{\{m_{\tilde e}\}_{\tilde e\in\partial i\setminus e}}}\quad\mathbb I\left(m+\sum_{\mathclap{\tilde e\in\partial i\setminus e}}m_{\tilde e}=\kappa_i\right)\ \prod_{\mathclap{\substack{\tilde e=(r,i)\\\tilde e\in\partial i\setminus e}}}\nu_{r\to \tilde e}(m_{\tilde e})\e^{-\beta m_{\tilde e}\omega_{\tilde e}} \ .
\end{equation}
We adopt the convention $\sum_{a\in A}f(a)=0$ and $\prod_{a\in A}f(a)=1$ if $A=\emptyset$ for any function $f$. {Pictorially, Eq.~\eqref{messaggio} can be rendered as 
\begin{center}
\newdimen\nodeSize
\nodeSize=4mm
\newdimen\nodeDist
\nodeDist=6mm
\tikzset{position/.style args={#1:#2 from #3}{at=(#3.#1), anchor=#1+180, shift=(#1:#2)}}
\begin{tikzpicture}[square/.style={regular polygon,regular polygon sides=4}]
\node[fill=black,square,draw,inner sep=0.2pt] (0) at (0,0) {\small \color{white} $i$};
\node[shape=circle,draw=none, position=0:20mm from 0] (ef) {};
\node[shape=circle, position=160:{0.5\nodeDist} from 0,draw,inner sep=0.6pt] (2) {\tiny $ui$};
\node[fill=black,square,position=160:{4\nodeDist} from 0,draw,inner sep=0.5pt] (a2) {\small \color{white} $u$};
\node[fill=gray!60, shape=rectangle, position=60:{1.5\nodeDist} from 2, draw,inner sep=2pt] (A2) {};
\node[fill=black,square,position=200:{4\nodeDist} from 0,draw,inner sep=0.5pt] (a3) {\small \color{white} $v$};
\node[shape=circle, position=200:{0.5\nodeDist} from 0,draw,inner sep=0.6pt] (3) {\tiny $vi$};
\node[ fill=gray!60, shape=rectangle, position=-60:{1.5\nodeDist} from 3, draw,inner sep=2pt] (A3) {};
\path[thin,->,-stealth] (0) edge node [fill=white] {$\nu_{i\to (ij)}$} (ef);
\draw[thin] (0) -- (2);
\path[thin,->,-stealth] (A2) edge node [fill=white] {$\small \e^{-\beta\omega_{ui}}$} (2);
\path[thin,->,-stealth] (A3) edge node [fill=white] {$\small \e^{-\beta\omega_{vi}}$} (3);
\draw[thin] (0) -- (3);
\node[draw=none,position=120:{0.35*\nodeDist} from a2,inner sep=1pt] (a2a) {};
\node[draw=none,position=180:{0.35*\nodeDist} from a2,inner sep=1pt] (a2b) {};
\node[draw=none,position=240:{0.35*\nodeDist} from a2,inner sep=1pt] (a2c) {};
\node[draw=none,position=-120:{0.35*\nodeDist} from a3,inner sep=1pt] (a3a) {};
\node[draw=none,position=-180:{0.35*\nodeDist} from a3,inner sep=1pt] (a3b) {};
\draw[thin,gray] (a2a) -- (a2) -- (a2b) -- (a2) -- (a2c);
\draw[thin,gray] (a3a) -- (a3) -- (a3b);
\path[thin,->,-stealth] (a2) edge node [fill=white,sloped] {\small $\nu_{u\to (ui)}$} (2);
\path[thin,->,-stealth] (a3) edge node [fill=white,sloped] {\small $\nu_{v\to (vi)}$} (3);
\end{tikzpicture}
\end{center}
where the arrows indicate the directions of ``propagation'' of the messages.} Taking advantage of the binary nature of the variables $m_e$, it is convenient to parametrize the marginals in terms of ``cavity fields'' $h_{i\to e}$,
\begin{equation}
 \nu_{i\to e}(m)=\frac{\e^{\beta m h_{i\to e}}}{1+\e^{\beta h_{i\to e}}},
\end{equation}
so that an equation for the cavity fields $h_{i\to e}$ can be written as
 \begin{multline}\label{cavbeta}
 h_{i\to e}=-\frac{1}{\beta}\ln\frac{\nu_{i\to e}(0)}{\nu_{i\to e}(1)} 
  =\frac{1}{\beta}\ln\sum_{\mathclap{\{m_{\hat e}\}_{\partial i\setminus e}}}\ \ \mathbb I\left(\sum_{\hat e\in \partial i\setminus e}m_{\hat e}=\kappa_i-1\right)\prod_{{\substack{\hat e=(r,i)\\\hat e\in\partial i\setminus e}}}\e^{\beta m_{\hat e}(h_{r\to \hat e}-\omega_{\hat e})}\\
-\frac{1}{\beta}\ln\sum_{\mathclap{\{m_{\hat e}\}_{\partial i\setminus e}}}\ \ \mathbb I\left(\sum_{\hat e\in \partial i\setminus e}m_{\hat e}=\kappa_i\right)\prod_{{\substack{\hat e=(r,i)\\\hat e\in\partial i\setminus e}}}\e^{\beta m_{\hat e}(h_{r\to \hat e}-\omega_{\hat e})}.
 \end{multline}
Once the equations have been solved for all the fields on the graph edges, the marginal probability of the variable $m$ on the edge $e=(i,j)$ is given by
\begin{equation}\label{marginal}
 \nu_{e}(m)=\frac{\e^{\beta m(h_{i\to e}+h_{j\to e}-\omega_{e})}}{1+\e^{\beta (h_{i\to e}+h_{j\to e}-\omega_{e})}},
\end{equation}
i.e., $\nu_{e}(1)$ evaluated with $\beta=1$ is the probability that $e\in\mF_k^*$. The BP approximation to the symbol MAP estimator in \eqref{eq_sMAP} is obtained computing the messages, and then the marginal, with $\beta=1$, and then selecting the set
\begin{equation}
\begin{split}
\hat{\mF}_{k}^{\rm (s)}(\mG)\coloneqq& \left\{e\in\mE\colon \nu_e(1)>\frac{1}{2}\right\}\\
=&\left\{e=(i,j)\in\mE\colon h_{i\to e}+h_{j\to e} > \omega_{e}\right\}.
\end{split}
\label{eq_inclusion_rule}
\end{equation}
Observe that proceeding in this way the selected edge-set $\hat{\mF}_{k}^{\rm (s)}\cup \mF_{k}^{\rm (0)} $ is not an actual $k$-factor in general.

The block MAP estimator is obtained taking $\beta\to +\infty$ in the equations for the marginals: in this limit the measure in Eq.~\eqref{lbeta} concentrates on the configuration $\{m_e\}_e$ that maximizes the likelihood. For $\beta\to+\infty$, Eqs.~\eqref{cavbeta} simplify, and we obtain
\begin{equation}\label{cav1}
h_{i\to e}={\min}^{(\kappa_i)}\left[\left\{\omega_{\hat e}-h_{r\to \hat e}\right\}_{{\substack{\hat e=(r,i)\\\hat e\in \partial i\setminus e}}}\right],
\end{equation}
where $\min^{(r)}[A]$ outputs the $r$th smallest element of the set $A$. The block MAP estimator $\hat \mF_k^{\rm (b)}$ is found using the same criterion given in Eq.~\eqref{eq_inclusion_rule} upon convergence of the algorithm, so that the final estimator for $\mF_k^*$ is $\hat \mF_k^{\rm (b)}\cup \mF_k^{(0)}$.

\subsection{Recursive Distributional Equations}
In this Section we will study the average error on the considered ensemble by analysing the statistical properties of the solutions of the BP equations by the cavity method~\cite{Mezard2001}. We introduce the following random variables.
\begin{itemize}
 \item $\hat {\mathsf H}$ is a random variable that has the law of the cavity field $h_{i \to e}$ given that $e$ is a randomly chosen planted edge;
 \item ${\mathsf H}$ is a random variable that has the law of the cavity field $h_{i \to e}$ given that $e$ is a randomly chosen non-planted edge;
 \item $\hat {\Omega}$ is a random variable that has the law of the effective weight $\omega_e$ given that $e$ is a randomly chosen planted edge;
 \item ${\Omega}$ is a random variable that has the law of the effective weight $\omega_e$ given that $e$ is a randomly chosen non-planted edge.
\end{itemize}

In the hypothesis that the replica symmetric hypothesis holds (i.e., typical realizations of $\nu$ have no long-range correlations), then \eqref{cavbeta} translates into recursive distributional equations (RDEs) involving the introduced random variables ${\mathsf H}$ and $\hat {\mathsf H}$. To write down this set of RDEs, first observe that an endpoint $i$ of a planted edge $e$ is incident to $\mathsf Z$ non-planted edges, plus a set $\mathsf K'-1$ of other planted edges. The random variable $\mathsf K'$, however, is not simply distributed as $\mathsf K$, but instead as \cite{mezard2009information}
\begin{equation}
\Prob[\mathsf K'=\kappa]=\frac{\kappa\Prob[\mathsf K=\kappa]}{\mathbb E[\mathsf K]}=\binom{k}{\kappa}\frac{\kappa\hat\mu^\kappa(1-\hat\mu)^{k-\kappa}}{k\hat\mu}.
\end{equation}
This is because the planted subgraph, having $\hat\mu_kN\gg 1$ vertices, contains $\frac{1}{2}\hat\mu_kN\kappa\Prob[\mathsf K=\kappa]$ edges adjacent to a vertex of capacity $\kappa$, so that the probability of picking a planted edge that is adjacent to a vertex with capacity $\kappa$ is proportional to $\kappa\Prob[\mathsf K=\kappa]$. For the sake of brevity, here and in the following, given a random variable $\mathsf X$, we will denote by $\mathsf X'$ a random variable distributed as
\begin{equation}\label{eqedgeper}
\Prob[\mathsf X'=x]=\frac{x\Prob[\mathsf X=x]}{\mathbb E[\mathsf X]}.
\end{equation}

Similarly, if $e$ is a non-planted edge and $i$ is one of its endpoints, there will be $\mathsf K$ planted edge incident to $i$, and other $\mathsf Z'-1$ non-planted edges. Within the replica-symmetric assumption of independence of the incoming cavity fields, one thus obtains from Eq.~\eqref{cavbeta}:
\begin{subequations}\label{rdebeta}
\begin{align}
 \hat{\mathsf H}\stackrel{\rm d}{=}&-\frac{1}{\beta}\ln\frac{\sum_{\{m\},\{\tilde m\}}\mathbb I\left(\sum_{i=1}^{\mathsf K'-1}m_i+\sum_{j=1}^{\mathsf Z}\tilde m_j=\mathsf K'\right)\prod_{i=1}^{\mathsf K'-1}\e^{\beta m_i(\hat{\mathsf H}_i-\hat\Omega_i)}\prod_{j=1}^{\mathsf Z}\e^{\beta \tilde m_j({\mathsf H}_j-\Omega_j)}}{\sum_{\{m\},\{\tilde m\}}\mathbb I\left(\sum_{i=1}^{\mathsf K'-1}m_i+\sum_{j=1}^{\mathsf Z}\tilde m_j=\mathsf K'-1\right)\prod_{i=1}^{\mathsf K'-1}\e^{\beta m_i(\hat{\mathsf H}_i-\hat\Omega_i)}\prod_{j=1}^{\mathsf Z}\e^{\beta \tilde m_j({\mathsf H}_j-\Omega_j)}}, \label{cavbetaH1}\\
{\mathsf H}\stackrel{\rm d}{=}&-\frac{1}{\beta}\ln\frac{\sum_{\{m\},\{\tilde m\}}\mathbb I\left(\sum_{i=1}^{\mathsf K}m_i+\sum_{j=1}^{\mathsf Z'-1}\tilde m_j=\mathsf K\right)\prod_{i=1}^{\mathsf K}\e^{\beta m_i(\hat{\mathsf H}_i-\hat\Omega_i)}\prod_{j=1}^{\mathsf Z'-1}\e^{\beta \tilde m_j({\mathsf H}_j-\Omega_j)}}{\sum_{\{m\},\{\tilde m\}}\mathbb I\left(\sum_{i=1}^{\mathsf K}m_i+\sum_{j=1}^{\mathsf Z'-1}\tilde m_j=\mathsf K-1\right)\prod_{i=1}^{\mathsf K}\e^{\beta m_i(\hat{\mathsf H}_i-\hat\Omega_i)}\prod_{j=1}^{\mathsf Z'-1}\e^{\beta \tilde m_j({\mathsf H}_j-\Omega_j)}}.\label{cavbetaH2}
\end{align}
\end{subequations}
The equalities have to be considered in distribution. In the equations above all random variables are independent, $\mathsf Z$ is Poisson distributed with mean $\gamma$, the $\Omega_i$'s have the same law as $\Omega$, ${\mathsf H}_i$ are independent copies of $\mathsf H$, and, similarly, $\hat {\mathsf H}_i$ of $\hat {\mathsf H}$. Finally, the variable $\mathsf K$ is distributed as in Eq.~\eqref{distK}. Observe that being $\mathsf Z$ Poisson distributed, $\mathsf Z'-1\stackrel{\rm d}{=}\mathsf Z$. Note also that, in the $\beta\to+\infty$ limit, Eqs.~\eqref{rdebeta} simplify, giving
\begin{subequations}\label{cavmap}
\begin{align}
 \hat{\mathsf H}\stackrel{\rm d}{=}&{\min_{ij}}^{(\mathsf K')}\left\{\{\hat\Omega_i-\hat{\mathsf H}_i\}_{i=1}^{\mathsf K'-1}\cup\{\Omega_i-\mathsf H_j\}_{j=1}^{\mathsf Z}\right\},\\
 {\mathsf H}\stackrel{\rm d}{=}&{\min_{ij}}^{(\mathsf K)}\left\{\{\hat\Omega_i-\hat{\mathsf H}_i\}_{i=1}^{\mathsf K}\cup\{\Omega_i-\mathsf H_j\}_{j=1}^{\mathsf Z'-1}\right\}.
\end{align}
\end{subequations}

Recalling the inclusion rule \eqref{eq_inclusion_rule}, the average reconstruction error is given as
\begin{equation}
\mathbb{E}[\varrho] = \frac{\hat\mu}{2} \Prob[\hat {\mathsf H}_1 + \hat {\mathsf H}_2 \le \hat \Omega  ]
+ \frac{\gamma \hat\mu_k}{2k} \Prob[\mathsf H_1 + \mathsf H_2 > \Omega  ].
\label{eq_averho_1}
\end{equation}

\subsection{Hard fields}
\label{sec:pruning2}
At this point, we aim at evaluating $\mathbb E[\varrho]$ by solving, possibly numerically, the RDEs given in Eqs.~\eqref{rdebeta}. It is, however, convenient to first isolate the contribution of ``hard-fields''. It is indeed not difficult to see that the events $\hat {\mathsf H} = + \infty$ and ${\mathsf H} = - \infty$ have finite probability. This follows from the fact that $\Prob[\mathsf Z=0]>0$ in \eqref{cavbetaH1}, which leads to $\hat {\mathsf H} = + \infty$, and this event can lead to ${\mathsf H} = - \infty$ in \eqref{cavbetaH2}. Let therefore be $\hat q\coloneqq\Prob[\hat{\mathsf H}<+\infty]$ and $q\coloneqq\Prob[{\mathsf H}>-\infty]$, probability that the fields are finite. We introduce two new random variables $\hat H$ and $H$ that have the law of $\hat {\mathsf H}$ and ${\mathsf H}$ conditional on being finite:
\begin{subequations}
\begin{align}
\mathsf H & \stackrel{\mathrm d}{=}\begin{cases} -\infty&\text{with prob. }1-q \ ,\\
           H&\text{with prob. } q \ ,
          \end{cases}
\\
\hat{\mathsf H} & \stackrel{\mathrm d}{=}\begin{cases} +\infty&\text{with prob. }1-\hat q \ ,\\
           \hat H&\text{with prob. } \hat q \ .
          \end{cases}
\end{align}
\end{subequations}
To obtain the equations obeyed by $q$, $\hat q$, $H$ and $\hat H$, it is convenient, in this sense, to observe that
\begin{subequations}
\begin{align}
 \hat{\mathsf H}\stackrel{\rm d}{=}&{\min}^{(\mathsf K')}\left\{\{\hat\Omega_i-\hat{\mathsf H}_i\}_{i=1}^{\mathsf K'-1}\cup\{\Omega_i-\mathsf H_j\}_{j=1}^{\mathsf Z}\right\}+O(\sfrac{1}{\beta}),\\
 {\mathsf H}\stackrel{\rm d}{=}&{\min}^{(\mathsf K)}\left\{\{\hat\Omega_i-\hat{\mathsf H}_i\}_{i=1}^{\mathsf K}\cup\{\Omega_i-\mathsf H_j\}_{j=1}^{\mathsf Z}\right\}+O(\sfrac{1}{\beta}),
\end{align}
\end{subequations}
the correction terms being finite. From these equations we easily get that 
\begin{subequations}\label{eqQQ}
\begin{align}
1-\hat q=&\e^{-\gamma q},\\
1-q=&\sum_{\kappa=1}^k\Prob[\mathsf K=\kappa](1-\hat q)^\kappa=1-\frac{1-(1-\hat\mu\hat q)^k}{1-(1-\hat\mu)^k},
\end{align}
\end{subequations}
that is a set of equations for $q$ and $\hat q$. Introducing the variables $Z$ distributed as
\begin{subequations}
\begin{equation}
\Prob[Z=z]=\frac{1-\hat q}{\hat q}\frac{(\gamma q)^z}{z!},\quad z\in\mathds N
\end{equation}
and the variable $K$ distributed as
\begin{equation}
\Prob[K=\kappa]=\binom{k}{\kappa}\frac{(1-\hat\mu \hat q)^{k-\kappa}(\hat\mu\hat q)^{\kappa}}{1-(1-\hat\mu \hat q)^k},\quad \kappa=1,\dots,k,
\end{equation}
\end{subequations}
we can reduce ourselves to equations involving `soft fields' $H$ and $\hat H$ only. Using the notation introduced in Eq.~\eqref{eqedgeper} we can write
\begin{subequations}\label{rdebetasoft}
\begin{align}
 \hat{ H}\stackrel{\rm d}{=}&-\frac{1}{\beta}\ln\frac{\sum_{\{m\},\{\tilde m\}}\mathbb I\left(\sum_{i=1}^{ K'-1}m_i+\sum_{j=1}^{ Z}\tilde m_j= K'\right)\prod_{i=1}^{ K'-1}\e^{\beta m_i(\hat{ H}_i-\hat\Omega_i)}\prod_{j=1}^{ Z}\e^{\beta \tilde m_j({ H}_j-\Omega_j)}}{\sum_{\{m\},\{\tilde m\}}\mathbb I\left(\sum_{i=1}^{ K'-1}m_i+\sum_{j=1}^{ Z}\tilde m_j= K'-1\right)\prod_{i=1}^{ K'-1}\e^{\beta m_i(\hat{ H}_i-\hat\Omega_i)}\prod_{j=1}^{ Z}\e^{\beta \tilde m_j({ H}_j-\Omega_j)}}, \label{cavbetaH1s}\\
{ H}\stackrel{\rm d}{=}&-\frac{1}{\beta}\ln\frac{\sum_{\{m\},\{\tilde m\}}\mathbb I\left(\sum_{i=1}^{ K}m_i+\sum_{j=1}^{ Z'-1}\tilde m_j= K\right)\prod_{i=1}^{ K}\e^{\beta m_i(\hat{ H}_i-\hat\Omega_i)}\prod_{j=1}^{ Z'-1}\e^{\beta \tilde m_j({ H}_j-\Omega_j)}}{\sum_{\{m\},\{\tilde m\}}\mathbb I\left(\sum_{i=1}^{ K}m_i+\sum_{j=1}^{ Z'-1}\tilde m_j= K-1\right)\prod_{i=1}^{ K}\e^{\beta m_i(\hat{ H}_i-\hat\Omega_i)}\prod_{j=1}^{ Z'-1}\e^{\beta \tilde m_j({ H}_j-\Omega_j)}}.\label{cavbetaH2s}
\end{align}
\end{subequations}
In the $\beta\to+\infty$ limit,
\begin{subequations}\label{cavmapsoft}
\begin{align}
 \hat{H}\stackrel{\rm d}{=}&{\min_{ij}}^{(K')}\left\{\{\hat\Omega_i-\hat{H}_i\}_{i=1}^{K'-1}\cup\{\Omega_i-H_j\}_{j=1}^{Z}\right\}\label{cavmapsoft1},\\
 {H}\stackrel{\rm d}{=}&{\min_{ij}}^{(K)}\left\{\{\hat\Omega_i-\hat{H}_i\}_{i=1}^{K}\cup\{\Omega_j-H_j\}_{j=1}^{Z'-1}\right\}.\label{cavmapsoft2}
\end{align}
\end{subequations}
It is important to note that the solution $\hat H=+\infty$ and $H=-\infty$ is an admissible solution of the RDEs for any value of $\beta$. This solution corresponds to a full recovery phase, in which the fraction of planted edges that are not correctly recovered vanishes as the system size grows.

The average reconstruction error \eqref{eq_averho_1} can be rewritten as
\begin{equation}
\mathbb{E}[\varrho] = \frac{\hat\mu \hat q^2}{2} \Prob[\hat  H_1 + \hat H_2 \le \hat \Omega  ] 
 + \frac{\hat\mu_k q^2\gamma }{2k} \Prob[H_1 + H_2\geq \Omega  ].
\label{eq_averho_2}
\end{equation}

This procedure of ``hard-fields'' elimination on the RDEs admits also an interpretation on a single graph instance. Infinite fields on the planted edges may appear in the BP equations \eqref{cavbeta}: if a vertex $i$ has coordination equal to its capacity $\kappa_i$, then $h_{i \to e}= +\infty$ for all its incident edges $e$. This is not surprising: in this case, indeed, all edges incident to $i$ surely belong to the planted $k$-factor. The vertex $i$ and all the $\kappa_i$ edges incident to it can be removed and the capacity of its neighbours updated. This removal procedure can be iterated until either $\mF_k^*$ has been entirely recovered, or a non-trivial core survives in which every vertex $i$ has $|\partial i|>\kappa_i$. The BP algorithm in Eq.~\eqref{cavbeta} can be runned on the obtained core.

The described pruning appears as a generalization of the known pruning procedure adopted for the study of optimal matching on Erd\H os-R\'enyi graphs \cite{Karp1981} and planted matching problems \cite{Semerjian2020}. It corresponds to a process of identification of the planted edges merely based on the topology of the graph $\mG$ (and therefore not related to the weights on the edges). A `percolation transition' occurs in the parameter of the problem between a phase in which the graph is completely pruned (and the $k$-factor completely recovered), and a phase in which an (extensive) core survives. The threshold is obtained studying the equation
\begin{equation}\label{eq:qeq}
q+\frac{\left(1-\hat\mu+\hat\mu\e^{-\gamma q}\right)^k-1}{\hat\mu_k}=0
\end{equation}
that has $q=0$ as attractor for
\begin{equation}\label{boundfr}
ck\mu\hat\mu\leq 1.
\end{equation}
In the considered problem, this condition implies full recovery of the planted configuration by simple topological considerations, i.e., iterative pruning.

\section{A numerical experiment: the exponential distribution case}
\label{sec:Exp}
We shall now numerically investigate the planted $k$-factor problem on the ensemble of graphs introduced above, using the tools described in the previous Section. We will take an uniform distribution for the non-planted weights,
\begin{equation}\label{uniform}
p(w)=\frac{1}{c}\mathbb I(0 \le w \le c),
\end{equation}
and an exponential distribution
\begin{equation}\label{expdis}
    \hat p(w)=\lambda \e^{-\lambda w}\mathbb{I}(w \ge 0) \ ,
\end{equation}
for the planted weights. It follows that in this case $\Gamma=[0,c]$, $\mu=1$, $\hat\mu=1-\e^{-c\lambda}$ and $\gamma=c(1-\e^{-k c\lambda})$. In the $c\to+\infty$ limit, $\hat\mu=\mu=1$ and $q=\hat q=1$ $\forall k\in\mathds N$.

\begin{figure*}
\subfloat[\label{fig:exprho}]{\includegraphics[height=0.45\columnwidth]{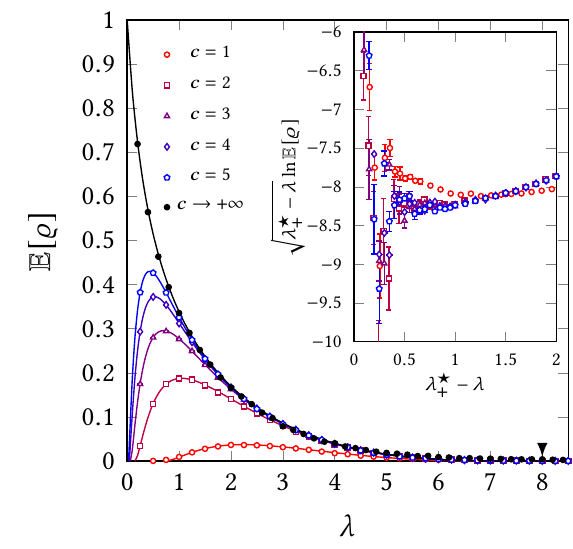}}\hfill
\subfloat[\label{fig:exprho1}]{\includegraphics[height=0.45\columnwidth]{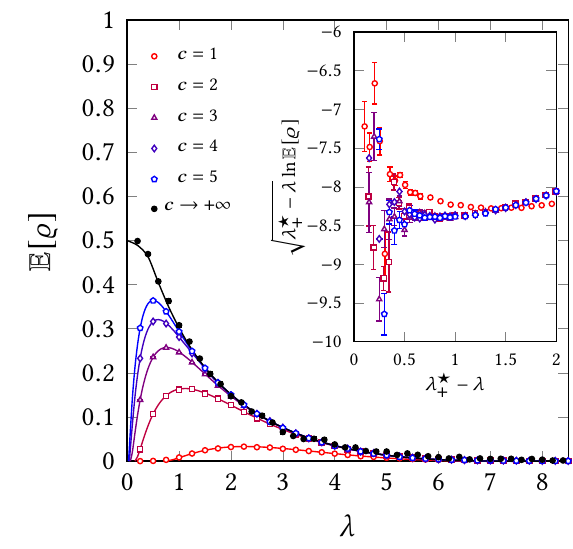}}\\\
\subfloat[\label{fig:drift}]{\includegraphics[height=0.45\columnwidth]{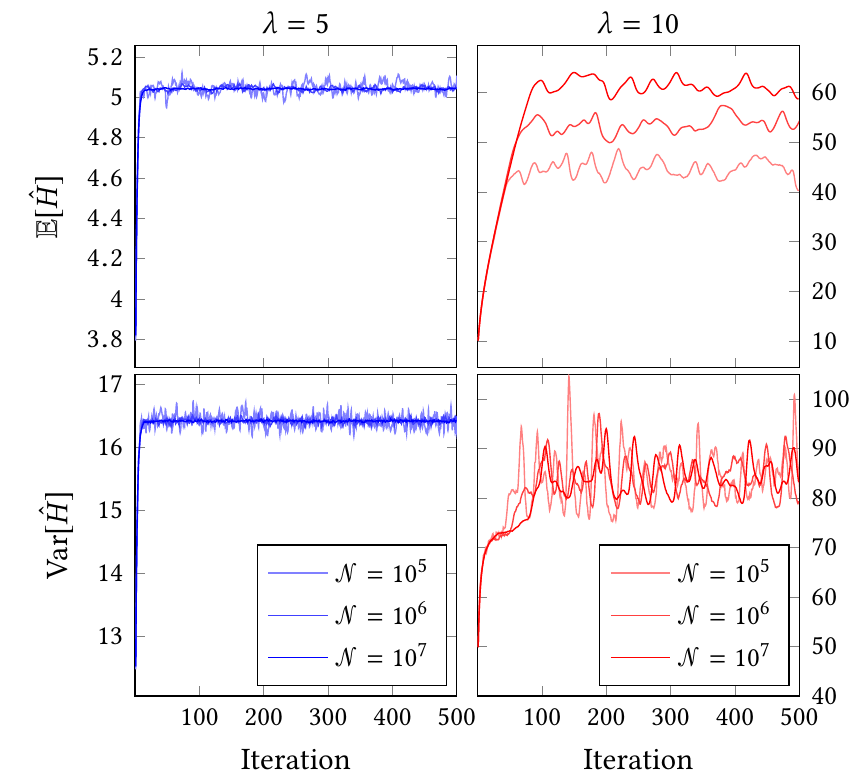}}\hfill
\subfloat[\label{fig:lambdap}]{\includegraphics[height=0.45\columnwidth]{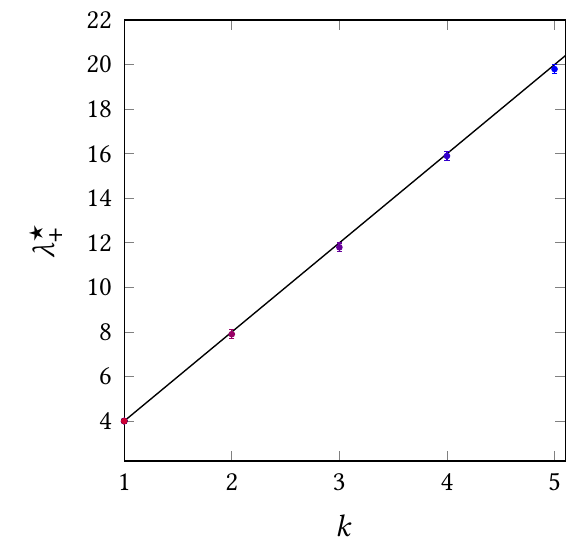}}
\caption{(a -- b) Reconstruction error using the block MAP (a) and the symbol MAP (b) estimators for the planted $2$-factor problem with exponential planted weights and different values of average degree $c$ of the non-planted edges. The lines have been obtained numerically solving the RDEs \eqref{cavmapsoft}, corresponding to $\beta=+\infty$, and \eqref{rdebetasoft} with $\beta=1$ respectively, using a PD algorithm with $\mathcal N=10^6$ fields. The dots are obtained solving $10^3$ instances of the problem using the BP algorithm in Eqs.~\eqref{cav1} and \eqref{cavbeta} with $\beta=1$ on graphs of $N=10^3$ vertices.  The $c\to+\infty$ curves are estimated using PD with $c=20$ and are compared with BP results obtained solving the problem on complete graphs with $N=10^2$ vertices. In the block MAP case, we marked with an arrow the partial to full recovery threshold in the $c\to+\infty$ limit predicted to be at $\lambda = 4k$. \textit{In the insets}, plot of $\sqrt{\lambda^\star_+-\lambda}\ln\mathbb E[\varrho]$ given by the PD algorithm, with $\lambda_+^\star$ estimated using the condition in Eq.~\eqref{condizione}. Both in the block MAP case and in the symbol MAP case, the limiting value for $\lambda\to\lambda^\star_+$ is finite, suggesting that the transition is of infinite order. (c) Mean and variance of the variable $\hat H$ estimated using a PD algorithm for $k=c=2$ for $\beta\to+\infty$. In this case, we numerically find $\lambda^\star_+=7.9(1)$, whereas the prediction given by Eq.~\eqref{condizione} is $\lambda^\star_+=7.9946\dots$. The results are obtained assuming all the $\mathcal N$ fields equal to zero as initial condition. An ``iteration'' of the algorithm corresponds to an update of all fields of the population by means of the RDEs. For $\lambda=5$, inside the partial recovery phase, the algorithm rapidly converges to asymptotic values that do not depend on the size $\mathcal N$ of the population. For $\lambda=10$, i.e., in the full recovery phase, the algorithm converges to values of $\mathbb E[\hat H]$ that are $\mathcal N$-dependent and diverge with $\mathcal N$, whereas the variance of the distribution goes to an $\mathcal N$-independent value (although noisy).  The described phenomenology appeared for all the investigated values of $c$ and $k$, and for both the block MAP and the symbol MAP. (d) Transition point $\lambda^\star_+$ for the block MAP in the exponential model for $c\to+\infty$. The continuous line correspond to the prediction $\lambda^\star_+=4k$ given in Section~\ref{sec:exp2}, whereas the dots are the numerical estimation of the transition point obtained using PD with $c=50$ and a population of $\mathcal N=10^8$ fields.}
\end{figure*}

The large $N$ limit results for $\mathbb E[\varrho]$ have been obtained by a numerical resolution of the RDEs for the soft fields for $\beta=1$ and $\beta\to+\infty$ (see Eqs.~\eqref{cavmapsoft}) via a population dynamics (PD) algorithm~\cite{Mezard2001}. The approach consists in introducing an iterative version of the RDEs discussed above that defines a new set of random variables $(\hat H^{(n)},H^{(n)})_n$, $n=0,1,\dots$. These new random variables satisfy, for $\beta\to+\infty$,
\begin{subequations}\label{cavmapsoftn}
\begin{align}
 \hat{H}^{(n+1)}\stackrel{\rm d}{=}&{\min_{ij}}^{(K')}\left\{\{\hat\Omega_i-\hat{H}^{(n)}_i\}_{i=1}^{K'-1}\cup\{\Omega_i-H_j^{(n)}\}_{j=1}^{Z}\right\},\\
 {H}^{(n)}\stackrel{\rm d}{=}&{\min_{ij}}^{(K)}\left\{\{\hat\Omega_i-\hat{H}^{(n)}_i\}_{i=1}^{K}\cup\{\Omega_j-H_j^{(n-1)}\}_{j=1}^{Z'-1}\right\},
\end{align}
\end{subequations}
with initial condition $\hat H^{(0)}=H^{(-1)}=0$. A similar set of iterative RDEs can be written for $\beta=1$ starting from Eqs.~\eqref{rdebetasoft}. The underlying assumption is that, if a non-trival solution of the original RDEs exists, such solution will be reached in probability by Eqs.~\eqref{cavmapsoftn} for $n\to+\infty$. 

From the algorithmic point of view, the law of the random variable $H^{(n)}$ is represented by an empirical distribution of a sample $\{h_1,\dots,h_{\mathcal N}\}$ of its representants, with $\mathcal N \gg 1$, so that
\begin{equation}
\Prob[H^{(n)}\leq h]\approx \frac{1}{\mathcal N} \sum_{i=1}^{\mathcal N} \mathbb{I}(h_i^{(n)} \le h).
\label{eq_population}
\end{equation}
Similarly, an empirical distribution is adopted to approximate the law of $\hat H^{(n)}$. The RDEs \eqref{cavmapsoftn} are used to update the population representing $(\hat H^{(n)},H^{(n)})$ to a new population representing the variables $(\hat H^{(n+1)},H^{(n+1)})$, each update corresponding to an ``iteration'' of the algorthm. The algorithm stops when some convergence criterion (usually the convergence of the moments of the populations) is satisfied (see, e.g., Ref.~\cite{mezard2009information} for additional details).

The PD predictions have been compared with actual BP results, obtained solving the inference problem on a large number of instances (see also Section~\ref{sec:app} for further details about the BP implementation).

In the block MAP case ($\beta\to+\infty$) it is observed that, for a given pair $k$ and $c$, there exists an interval $\Lambda^{(\rm b)}\coloneqq (\lambda^\star_{-},\lambda^\star_{+})$ of values of $\lambda$, such that for $\lambda\in\Lambda^{(\rm b)}$ the PD algorithm converges to a finite solution. The corresponding value of $\mathbb E[\varrho]$ is found to be non-zero and, within numerical precision, $\mathbb E[\varrho]\to 0$ smoothly on the boundary of the interval. In particular, we numerically observe that
\begin{equation}\lim_{\lambda\to\lambda^\star_+}\sqrt{\lambda^\star_+-\lambda}\ln\mathbb E[\varrho]=-\alpha,\qquad \alpha>0.\end{equation}
This implies that $\mathbb E[\varrho]$ approaches zero exponentially fast as $\lambda\to\lambda^\star_+$, i.e., the transition is of infinite order. Remarkably, the very same behavior has been observed in the $k=1$ and $c\to+\infty$ case \cite{Semerjian2020}, where it is shown that $\lambda_+^\star=4$ and
\begin{equation}\ln\mathbb E[\varrho]=-\frac{2\pi}{\sqrt{4-\lambda}}-\frac{3}{2}\ln(4-\lambda)+O(1)\end{equation}
Finally, for $c\to +\infty$ it is observed that $\lambda^\star_{-}\to 0$, whereas $\lambda^\star_{+}$ approaches a finite limit. The obtained prediction for $\mathbb E[\varrho]$ is fully compatible with the BP results, and $\mathbb E[\varrho]\to 0$ for $\lambda\to\lambda^\star_\pm$ with the size of the considered graphs.

On the other hand, for $\lambda\not\in\Lambda^{(\rm b)}$ the PD algorithm does not converge. To be more precise, the population is subject to a `drift' towards the full recovery solution, $H\to -\infty$ and $\hat H\to+\infty$. This can be seen, for example, in Fig.~\ref{fig:drift}, where some numerical results for $c=k=2$ are given in the $\beta\to+\infty$ case for $\hat H$. It is seen that the numerically estimated mean $\mathbb E[\hat H]$ is population-size-dependent, and in particular diverges with the population size, whereas the variance is not. Moreover, larger populations correspond to larger values of $\mathbb E[\hat H]$.

The numerical results suggest therefore that a nontrivial, attractive fixed point exists for $\lambda\in\Lambda^{(\rm b)}$ only, otherwise the only attractor being the trivial fixed point $\hat H=-H=+\infty$ corresponding to the full recovery phase. In Ref.~\cite{Semerjian2020} it is argued that, for $k=1$, an infinite order phase transition takes place between a full recovery phase and a partial recovery phase, and in particular full recovery is obtained for $\lambda\in\mathds R^+\setminus\Lambda^{(\rm b)}$. The conjectures in Ref.~\cite{Semerjian2020} about the location of the transition and its nature have been recently rigorously proved in Ref.~\cite{Ding2021}. Our results strongly suggest that the same phenomenology extends to the $k>1$ case. As in the $k=1$ case, the accurate numerical determination of the endpoints of $\Lambda^{(\rm b)}$ is heavily affected by finite-population-size effects using PD (and finite-size effects using BP). Indeed, the transition manifests itself as a front propagation in the cumulative distribution function that drifts towards large values of the fields. Such front propagations are generically driven by the behavior in the exponentially small tail far away from the front~\cite{Brunet1997}. The finite population size induces a cutoff on the smallest representable value of the cumulative distribution function, that translates, assuming an exponential decay of the cumulative, into logarithmic finite population size effects on the velocity of the front and the location of the transition.

The very same phenomenology is observed for $\beta=1$, i.e., for the symbol MAP, where a partial recovery phase $\lambda\in\Lambda^{(\rm s)}=(\lambda^\star_-,\lambda^\star_+)$ is surrounded by a full recovery phase. For a given pair $k$ and $c$ of parameters, the symbol MAP transition points are found to be very close to the block MAP transition points obtained for the same values of $k$ and $c$. Also in this case, it is found that $(\lambda^\star_+-\lambda)^{\sfrac{1}{2}}\ln\mathbb E[\varrho]\to -\hat\alpha$ for $\lambda\to \lambda^\star_+$ for some $\hat\alpha>0$, suggesting that the transition between the partial recovery phase and the full recovery phase is of infinite order also in this case. In Fig.~\ref{fig:exprho1} we give the PD results for $\mathbb E[\varrho]$ in this case, alongside with the results of the BP simulations.

\section{A criterion for the block MAP transition}\label{sec:criterion}
In this Section we give a heuristic criterion for the transition between the partial recovery phase $\mathbb E[\varrho]>0$, and the full recovery phase $\mathbb E[\varrho]=0$ in the case of the block MAP. Our reasoning will follow and generalize the one given in Ref.~\cite{Semerjian2020} for the $k=1$ case. Our approach is inspired by the physics literature on front propagation in reaction-diffusion systems and equations of the FKPP type \cite{Brunet1997,Majumdar2000,Ebert2000}. Before applying it, however, an additional simplification of Eqs~\eqref{cavmapsoft} must be performed. We will proceed in generality, assuming that $\hat p$ depends on a parameter, let us call it $\lambda$. Moreover, we will assume that a special value $\lambda^\star$ exists such that for $\lambda<\lambda^\star$ we are in a partial-recovery phase, whereas for $\lambda>\lambda^\star$ we are in a full-recovery phase. We will also assume that the transition is \textit{continuous}, i.e., $\mathbb E[\varrho]\to 0$ smoothly as $\lambda\to {\lambda^\star}^-$. Observe that $\mathbb E[\varrho]\to 0$ means that $\Pr[H_1+H_2<\Omega]\to 1$ and $\Pr[\hat H_1+\hat H_2>\hat\Omega]\to 1$, see Eq.~\eqref{eq_averho_2}. The first property implies that, approaching the transition, $H_1<\Omega-H_2$ almost surely, i.e., in Eq.~\eqref{cavmapsoft2} the minimum picks almost surely one of the `planted contributions'. Similarly, the second property implies that in the same limit the minimum in Eq.~\eqref{cavmapsoft1} is almost surely picked in the set of `non-planted contributions'. These observations lead us to introduce a new set of random variables $(\hat U^{(n)},U^{(n)})_n$, satisfying the iterative RDEs,
\begin{align}\label{eqK}
 \hat{U}^{(n+1)}\stackrel{\rm d}{=}&{\min_{1\leq i\leq Z}}\{\Omega_i-{U}^{(n)}_i\},\\
 {U}^{(n)}\stackrel{\rm d}{=}&{\max_{1\leq i\leq K}}\{\hat\Omega_i-\hat U^{(n)}_i\},
\end{align}
with initial condition $\hat U^{(0)}=U^{(0)}=0$, corresponding to the expected ``effective'' behavior of Eqs.~\eqref{cavmapsoftn} near the transition. The new set of auxiliary variables is informative on the behavior of the random variables $(\hat H^{(n)},H^{(n)})_n$. Indeed, we can prove that
\begin{equation}\label{ordering}
\hat U^{(n)}\preceq  \hat H^{(n)},\qquad H^{(n-1)}\preceq  U^{(n)},\qquad \forall n.
\end{equation}
Given two random variables $X$ and $Y$, we say that $X\preceq Y$ if $\mathbb P[X>z]\leq \mathbb P[Y>z]$ for all $z$ \cite{Lindvall2012}. The proof proceeds by induction. Eqs.~\eqref{ordering} are satisfied for $n=0$. Assuming that they are satisfied for given $n$, it is easily proved that they are satisfied for $n+1$, being
\begin{subequations}
\begin{align}
H^{(n)}&\preceq\max_{1\leq i\leq K}\{\hat\Omega_i-\hat{H}_i^{(n-1)}\}\preceq \max_{1\leq i\leq K}\{\hat\Omega_i-\hat{U}_i^{(n)}\}={U}^{(n+1)},\\
\hat {U}^{(n+1)}&=\min_{1\leq j\leq Z}\{\Omega_j-{U}^{(n)}_j\}\preceq\min_{1\leq j\leq Z}\{\Omega_j-{H}^{(n)}_j\}\preceq {H}^{(n+1)}. 
\end{align}
\end{subequations}
The result stated above implies that, if $\Pr[\hat U^{(n)}>z]\to 1$ for $n\to+\infty$, then $\Pr[\hat H^{(n)}>z]\to 1$ as well in the same limit. We will obtain now a sufficient condition to have $\Pr[\hat U^{(n)}>z]\to 1$ that we will give us, therefore, a (sufficient) criterion to be in the full recovery phase. We define
\begin{align}
 F(x;n)&\coloneqq \Prob[\hat U^{(n)}<x],\\
\Phi(x;n)&\coloneqq\Prob[U^{(n)}<x].
\end{align}
Denoting by $\mathbb E_X[\bullet]$ the expectation with respect to the random variable $X$, we have that
\begin{align}
F(x;n+1)&=1-\mathbb E_Z\left[\left(\mathbb E_\Omega\left[\Phi(\Omega-x;n)\right]\right)^Z\right],\\
\Phi(x;n)&=\mathbb E_K\left[\left(1-\mathbb E_{\hat\Omega}\left[ F(\hat\Omega-x;n) \right]\right)^K\right],
\end{align}
and therefore
\begin{equation}\label{eqnonlinearizzata}
 F(x;n+1)=
 1-\mathbb E_Z\left[\left(\mathbb E_{\Omega}\left[\left(1-\mathbb E_{\hat\Omega}\left[F(\hat\Omega-\Omega+x;n)\right]\right)^K\right]\right)^Z\right].
\end{equation}
Suppose now that the cumulative $F(x;n)$ is subject to a `drift', i.e., there exists a velocity $v$ such that $F(x+vn;n)\to F(x)$ as $n\to\infty$. This is in line with the numerical result, which suggest that $\mathbb E[\hat H]\to +\infty$ and $\mathbb E[H]\to -\infty$ approaching the transition, whereas the higher order cumulants remain finite. Fig.~\ref{fig:drift}, in particular, shows that the means of the distributions are subject to a constant drift velocity that (at the leading order in $n$) does not depend on $n$. Moreover, this assumption is compatible with what has been observed in \cite{Semerjian2020}, and rigorously proved in \cite{Kingman1975,Biggins1977}, in the study of Eq.~\eqref{eqK} for $K\equiv 1$. Then, for $n\to+\infty$,
\begin{equation}
 F(x-v)=
 1-\mathbb E_Z\left[\left(\mathbb E_{\Omega}\left[\left(1-\mathbb E_{\hat\Omega}\left[F(\hat\Omega-\Omega+x)\right]\right)^K\right]\right)^Z\right].
\end{equation}
For $x\to-\infty$, $F(x)\to 0$ by definition, and in this limit at first order in $F$
\begin{equation}\label{linearizzata}
 F(x-v)\simeq \mathbb E[Z]\mathbb E[K]\mathbb E\left[ F(\hat\Omega-\Omega+x)\right]
\end{equation}
(we have dropped the subscripts implying an average over all variables in the argument). This linear (integral) equation has a solution in the form $F_v(z)=\e^{\theta z}$, with $\theta >0$ to respect the increasing character of distribution functions. Indeed, plugging this solution into the linearized equation we have
\begin{equation}
v(\theta)\simeq  -\frac{\ln\left(\mathbb E[Z]\mathbb E[K]\mathbb E\left[\exp(\theta\hat\Omega-\theta\Omega)\right]\right)}{\theta}.
\end{equation}
The choice of the appropriate $\theta$ to estimate the drift velocity is, at this point, not obvious. It can be shown  \cite{Kingman1975,Biggins1977,Brunet1997,Majumdar2000,Ebert2000} that the relevant value $\theta^*$ is the one that corresponds to the \textit{maximum} velocity, i.e., $\theta^*=\arg\sup_{\theta>0} v(\theta)$, and therefore
\begin{subequations}
\begin{align}\label{velocitas}
 v=&-\inf_{\theta>0}\frac{\ln\left(\mathbb E[Z]\mathbb E[K]\mathbb E\left[\exp(\theta\hat\Omega-\theta\Omega)\right]\right)}{\theta}\\
 =&-\inf_{\theta>0}\frac{\ln\left[\mathcal I(\theta)\mathcal I(1-\theta)\right]}{\theta}
\end{align}
\end{subequations}
where
\begin{equation}
\mathcal I(\theta)\coloneqq\sqrt{ck}\int_\Gamma \hat p^{\theta}(w)p^{1-\theta}(w)\dd w.
\end{equation}
If $v=v(\theta^*)>0$, then the distribution drifts towards $+\infty$ and we are in a full recovery phase ($\hat H\to +\infty$).  We postulate that the marginal condition $v(\theta^*)=0\Rightarrow \ln\left[\mathcal I(\theta^*)\mathcal I(1-\theta^*)\right]=0$ corresponds to the transition point. Being $\ln(\mathcal I(\theta)\mathcal I(1-\theta))$ a convex function symmetric around $\theta=\sfrac{1}{2}$, one has $v = 0$ when $\mathcal I(\sfrac{1}{2}) = 1$. This condition can be written as
\begin{equation}\label{condizione2}
\frac{{\rm D}_{\sfrac{1}{2}}(\hat p\|p)}{\ln(ck)}=1,
\end{equation} 
where
\begin{equation}
{\rm D}_{\alpha}(p\|q)\coloneqq \frac{1}{\alpha-1}\ln\int p^\alpha(x)q^{1-\alpha}(x)\dd x,
\end{equation}
is the R\'enyi divergence of order $\alpha$. In an equivalent form, Eq.~\eqref{condizione2} can be written as
\begin{equation}\label{condizione}
\int_\Gamma\sqrt{\hat p(w)p(w)}\dd w
=\frac{1}{\sqrt{c k}}.
\end{equation}
The condition above generalizes the one obtained for the planted matching problem \cite{Semerjian2020}, that is recovered for $k=1$. 

As a final comment, observe that, being by construction $\mathbb E[\varrho(\hat \mF_k^{(\rm s)},\mF^*_k)]\leq \mathbb E[\varrho(\hat \mF_k^{(\rm b)},\mF^*_k)]$, full recovery by means of the block MAP implies full recovery by means of the symbol MAP, and the partial recovery interval obtained using the symbol MAP is contained in the partial recovery interval of the block MAP.

\begin{figure}{\begin{center}
\includegraphics[width=0.5\columnwidth]{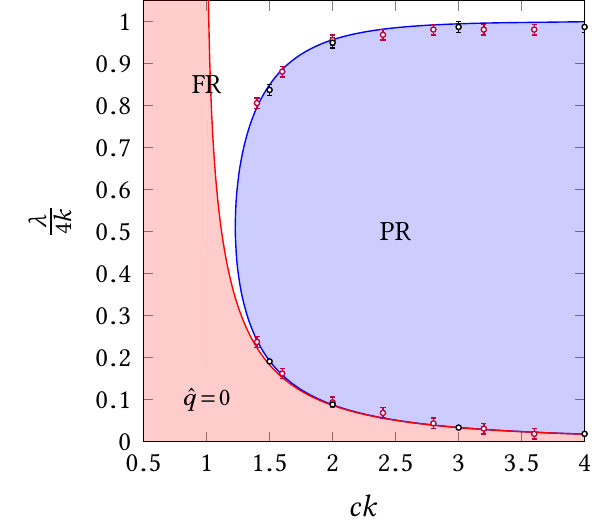}\end{center}}
\caption{Phase diagram of the planted $k$-factor with exponential planted weights on sparse graphs. Our argument in Section~\ref{sec:criterion} predicts a partial recovery (PR) phase (in blue) and a full recovery phase, corresponding to the remaining portion of the plane, depending on the values of $\lambda$ and $c$. Within the full recovery phase, the red area corresponds to the set of parameters for which full recovery is possible by means of pruning, i.e., $\hat q=q=0$. The red dots have been obtained from the numerical resolution of the RDEs \eqref{cav1} for the $2$-factor by a population dynamics algorithm with $\mathcal N=10^7$. The black dots correspond instead to the $k=1$ case and are taken from Ref.~\cite{Semerjian2020}. \label{fig:phase}}
\end{figure}

\section{Examples of results}\label{sec:app}
In this Section, we consider some special formulations of the planted $k$-factor problem, and we compare our numerical results with the theoretical predictions obtained from the general criterion given in Section~\ref{sec:criterion}. 

The numerical results are obtained studying the RDEs \eqref{rdebeta} for the problem by means of a population dynamics algorithm for $\beta=1$ (symbol MAP) and $\beta\to+\infty$ (block MAP). 

We also implemented a BP algorithm for the solution of the problem on actual graphs by means of the algorithm in Eq.~\eqref{cavbeta}. In particular, a random weighted graph with $N$ vertices is generated according to the ensemble introduced in Section~\ref{sec:definitions}. This graph is first subject to the pruning procedures described in Section~\ref{sec:pruning} and Section~\ref{sec:pruning2}. Given an edge $e=(i,j)$ of the resulting graph, we associate two fields $h_{i\to e}$ and $h_{j\to e}$ to it, initialized to random values. The fields are updated using Eq.~\eqref{cavbeta} if $\beta>0$ or using Eqs.~\eqref{cav1} if $\beta=+\infty$ for a large number of iterations. A candidate solution $\hat\mF_k$ is then selected using the criterion in Eq.~\eqref{eq_inclusion_rule}. In all cases, we stopped the algorithm after $5N$ updates, or before if the set $\hat\mF_k$ does not change for at least $50$ iterations. The error $\varrho$ is then obtained using Eq.~\eqref{eq_def_rho}, and the average error $\mathbb E[\varrho]$ is estimated considering a large number of independent instances of the problem.

The BP algorithm for the estimation of the block MAP given in Eq.~\eqref{cav1} coincides with the BP algorithm for the minimum weight $k$-factor introduced by Bayati and coworkers \cite{Bayati2011}. They proved therein that the algorithm converges in polynomial time to the correct minimizer as long as there are no fractional solutions, i.e., solutions with non-integer values of $m_e$. A worst-case analysis of the convergence properties of the BP algorithm with $\beta=1$, on the other hand, is still missing. Observe that the BP algorithm has longer running time for $c\gg k$ when $\beta$ is finite with respect to the $\beta\to+\infty$ case. In this case, indeed, given a node with valence $\kappa$, each step in Eq.~\eqref{cavbeta} requires the sum of $O(c^\kappa)$ contributions. Eq.~\eqref{cav1}, instead, asks only for to the $\kappa$-th incoming field, an operation that requires $O(c)$ steps. On complete graphs, having $c=N-k-1$, this means that the algorithm running time is increased by a factor $N^{k-1}$ in the finite $\beta$ case with respect to the $\beta=+\infty$ case.

\subsection{The fully-connected case}
\label{sec:fully_connected}
Dense models are recovered in our setting considering the $c\to+\infty$ limit, to be taken after the $N\to+\infty$ limit. Assuming that $\mu\hat\mu_k\neq 0$ (the problem is otherwise trivial) this implies $\gamma\to+\infty$ and therefore $\hat q=q=1$ because of Eqs.~\eqref{eqQQ}, meaning that in the thermodynamic limit there are (almost surely) no hard fields.

Eq.~\eqref{condizione2} also implies that, if $\lim_{c\to+\infty}{\rm D}_{\sfrac{1}{2}}(\hat p\|p)<+\infty$ no transition can take place in the fully connected limit. To get nontrivial results in this limit, it is therefore necessary to scale the weights with $c$, so that at the transition
\begin{equation}\label{condizione4}
{\rm D}_{\sfrac{1}{2}}(\hat p\|p)=\ln c+o(\ln c)\quad\text{for}\quad c\gg 1. 
\end{equation}
Suppose, for example, that $p$ is $c$-independent, and $\hat p(w)\equiv c^{-a}f(wc^{-a};b)$ with $a>0$ and $b$ parameters, and some function $f$ such that $f_0(b)\coloneqq\lim_{x\to 0}f(x;b)\in(0,+\infty)$. Then the condition \eqref{condizione4} implies that the threshold for $c\to+\infty$ is at
\begin{equation}
a=1.
\end{equation}
If instead $\hat p(w)\equiv c^{-1}f(wc^{-1};b)$, then the asymptotic formula \eqref{condizione4}  is not sufficient anymore and Eq.~\eqref{condizione} must be considered. The threshold condition becomes
\begin{equation}
\sqrt{f_0(b)}\int\sqrt{p(w)}\dd w=\frac{1}{\sqrt{k}}.
\end{equation}

The observations above are compatible with the rigorous results obtained by Bagaria and coworkers for the planted $2$-factor problem on complete graphs of $N$ vertices for $N\to+\infty$ \cite{Bagaria2018}. In their paper, they prove that, for $k=2$, on the threshold the following limit holds
\begin{equation}\label{eq:Bagaria}
\liminf_{N\to +\infty}\frac{{\rm D}_{\sfrac{1}{2}}(\hat p\|p)}{\ln N}=1 
\end{equation}
under some assumptions on the distributions $p$ and $\hat p$ (fulfilled, e.g., by Gaussian or exponential distributions, see Ref.~\cite{Bagaria2018} for details). This condition corresponds to Eq.~\eqref{condizione4}, observing that on a complete graph $c=N-k-1$. As we will show below, however, Eq.~\eqref{eq:Bagaria} can be not sufficient to recover the transition point.

\subsection{The exponential model}\label{sec:exp2}
Let us now briefly revisit the exponential case discussed in Section~\ref{sec:Exp}. If we apply Eq.~\eqref{condizione} to derive the block MAP threshold, we obtain
\begin{equation}
2-2\e^{-\frac{c\lambda}{2}}=\sqrt{\frac{\lambda}{k}}.
\end{equation}
Introducing $s=ck$ and $t=\lambda k^{-1}$, one parameter can be absorbed obtaining
\begin{equation}
2-2\e^{-\frac{st}{2}}=\sqrt{t}.
\end{equation}
This equation is always solved by $t=\lambda=0$. For $s=ck\gtrsim 1.2277\dots$, two additional solutions for $t$, and therefore $\lambda$, appear, let us call them $\lambda^\star_-$ and $\lambda^\star_+$, delimiting the partial recovery phase, see Fig.~\ref{fig:phase}. In the $c\to +\infty$ limit, $\lambda^\star_-\simeq\sfrac{1}{c}\to 0$ and only one transition point is found
\begin{equation}
\lambda^\star_+=4k. 
\end{equation}
This result is confirmed by the numerics, see Fig.~\ref{fig:lambdap}. For $k=1$ we recover the known result $\lambda^\star_+=4$, rigorously proved in \cite{Moharrami2019}. The criterion predicts that $4k-\lambda^\star_+$ approaches zero as $\e^{-ck}$ for $ck\to+\infty$. Observe that in this case, considering $c=N-k-1$, ${\rm D}_{\sfrac{1}{2}}(\hat p\|p)=\ln N+O(N)$ for all values of $k$ and $\lambda>0$. In other words, Eq.~\eqref{eq:Bagaria} of \cite{Bagaria2018}, although verified on the transition, is not enough to recover the threshold.

\subsection{Hidden Hamiltonian cycle recovery}\label{sec:HiddenHam}
In this section we are interested in solving a special type of planted $2$-factor problem, namely the hidden Hamiltonian cycle recovery (HC) problem. This is a planted $2$-factor problem in which the hidden $2$-factor is connected, i.e., it is a Hamiltonian cycle of the graph. The very same BP algorithms discussed for the planted $2$-factor can be applied to recover the hidden Hamiltonian cycle. 

\begin{figure}
{
\begin{center}
\includegraphics[width=0.7\columnwidth]{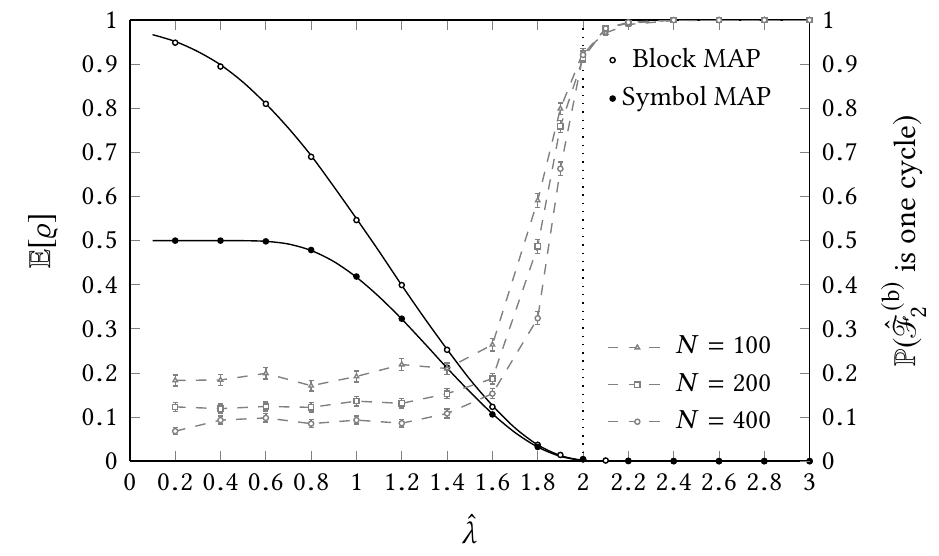}
\end{center}}
\caption{Recovery using the block MAP and the symbol MAP in the HC problem with $\hat p=\mathzapf N(\hat\lambda\sqrt{\ln N},1)$ and $p=\mathzapf N(0,1)$. The dots are obtained running BP on $10^3$ instances of complete graphs with $N=100$ with a hidden Hamiltonian cycle to be recovered. The black lines correspond to the PD prediction for the planted $2$-factor with $c=100$, with same planted and non-planted weight distributions. The vertical line indicates the transition point between partial and full recovery predicted by the theory for the block MAP. In gray, we plot the probability that the block MAP estimator $\hat\mF^{(\rm b)}_{k=2}$ is a Hamiltonian cycle (instead of a union or two or more cycles) for different sizes of the problem.\label{fig:phaseham}}
\end{figure}

This problem was studied in Ref.~\cite{Bagaria2018}, with the planted weights being assumed to be normal variables, $\hat p=\mathzapf N(\lambda,1)$, whereas the non-planted weights have distribution $p=\mathzapf N(0,1)$. In this case, therefore, ${\rm D}_{\sfrac{1}{2}}(\hat p\|p)=\frac{1}{4}\lambda^2$. Applying Eq.~\eqref{eq:Bagaria}, a nontrivial transition in the block MAP estimator is expected at $\lambda^2=4\ln N+o(\ln N)$. Parametrizing $\hat p$ with $\lambda^2=\hat\lambda^2\ln N$, the transition is then at $\hat\lambda=2$. This is rigorously proved and numerically verified in Ref.~\cite{Bagaria2018}, see Fig.~\ref{fig:phaseham}. In Fig.~\ref{fig:phaseham} we also plot, for the block MAP case, the probability that the estimator $\hat{\mathrsfso F}_2$ provided by the BP algorithm is actually connected, i.e., it is a single Hamiltonian cycle. Our numerics suggest that, for $\hat\lambda>2$, this probability goes to $1$ as $N\to+\infty$.

As discussed in Section~\ref{sec:definitions}, however, the block MAP is not the estimator that minimizes the error in Eq.~\eqref{eq_def_rho}. If we aim at minimizing the error $\varrho$, then the optimal estimator is the symbol MAP. We recall that this estimator does not provide a $k$-factor in general. Running our BP algorithm on graphs with hidden Hamiltonian cycles in the ensemble considered in \cite{Bagaria2018}, we obtained the results in Fig.~\ref{fig:phaseham}. The average error obtained using the symbol MAP is found to be smaller than the one obtained using the block MAP, as expected. On the other hand, finding the symbol MAP is computationally more expensive, as discussed above. 
For comparison, in Fig.~\ref{fig:phaseham} we plot also the PD results for the $2$-factor, finding a good agreement between the infinite-size prediction of the $2$-factor problem and the BP results of the HC problem also in the partial recovery phase.

\medskip

\section{Perspectives}
\label{sec:conclusions}

The transition appearing in the planted $k$-factor problem is of the same type as found in the planted matching problem \cite{Moharrami2019,Semerjian2020,Ding2021} and separates a partial recovery phase from a full recovery phase. Using heuristic arguments based on the literature on front propagation for reaction-diffusion equations, we have been able to obtain a simple and explicit criterion for the transition. We numerically tested the transition criterion, and we checked its consistency with the known results on the recovery thresholds of the planted $2$-factor problem. A rigorous proof of this transition criterion remains, however, as an open problem.

The heuristic argument is based on the fact (numerically observed) that the phase transition is continuous. It is not excluded a priori that first order transitions are possible for some nontrivial choice of the weight distributions or degree distributions of the graph, allowing the presence of multiple BP fixed points \cite{Bordenave2013}.

Finally, the threshold criterion obtained in the paper concerns the block MAP and only provides a bound for full recovery by means of the symbol MAP. In the numerically investigated cases, the recovery thresholds of the symbol MAP are observed to be very close to the ones of the block MAP. A formula for the exact location of the symbol MAP transition (and possibly its relation with the block MAP transition) is however still missing.

\subsection*{Acknowledgments}
The authors acknowledge collaboration with Guilhem Semerjian on the work \cite{Semerjian2020} that was the source of the key theoretical ideas underlying the analysis in the present paper and for many insightful discussions on the problem. This project has received funding from the European Union's Horizon 2020 research and innovation program under the Marie Sk\l{}odowska-Curie grant agreement CoSP No 823748, and from the French Agence Nationale de la Recherche under grant ANR-17-CE23-0023-01 PAIL.

\section*{References}
\bibliography{bibliografia}
\end{document}